\documentclass[%
 reprint,
nofootinbib,
 amsmath,amssymb,
 aps,
]{revtex4-1}

\newenvironment{exer*}
  {\ex}
  {\endex}

\def\be{\begin{equation}}
\def\ee{\end{equation}}
\def\bea{\begin{eqnarray}}
\def\eea{\end{eqnarray}}
\newcommand{\ket}[1]{\mbox{$|#1\rangle$}}
\newcommand{\bra}[1]{\mbox{$\langle#1|$}}

\newcommand{\m}[1]{\mbox{$\mathbf{#1}$}}

\usepackage{xspace} 

\newcommand{\hairsp}{\hspace{1pt}} 
\newcommand{\ie}{\textit{i.\hairsp{}e.}\xspace} 

\newcommand{\trm}[1]{\textrm{#1}}
\newcommand{\um}{\mu\trm{m}}
\newcommand{\mm}{\trm{mm}}
\newcommand{\nm}{\trm{nm}}

\newcommand{\prens}[1]{\left(#1\right)}
\newcommand{\vecb}[1]{\mathbf{#1}}

\newcommand{\perm}{\varepsilon}

\newcommand{\E}{\vecb{E}}

\newcommand{\alpham}{\alpha_\trm{m}}

\usepackage{graphicx}
\usepackage{pgfplots}
\usepackage{subfigure}
\usepackage{xcolor}
\usepackage{tikz}
\RequirePackage{atbegshi}
\usetikzlibrary{positioning}
\usetikzlibrary{arrows}
\usetikzlibrary{decorations.text}

\let\svthefootnote\thefootnote
\newcommand\blankfootnote[1]{%
  \let\thefootnote\relax\footnotetext{#1}%
  \let\thefootnote\svthefootnote%
}
\let\svfootnote\footnote
\renewcommand\footnote[2][?]{%
  \if\relax#1\relax%
    \blankfootnote{#2}%
  \else%
    \if?#1\svfootnote{#2}\else\svfootnote[#1]{#2}\fi%
  \fi
 }

\begin{document}

\preprint{APS/123-QED}

\title{Optomechanical antennas for on-chip beam-steering}

\author{Christopher Sarabalis$^{\dagger}$, Rapha\"{e}l Van Laer$^{\dagger}$ and Amir H. Safavi-Naeini}

\affiliation{Department of Applied Physics, and Ginzton Laboratory, Stanford University, Stanford, California 94305, USA$^{\star}$}

\date{\today}

\begin{abstract}
Rapid and low-power control over the direction of a radiating light field is a major challenge in photonics and a key enabling technology for emerging sensors and free-space communication links. Current approaches based on bulky motorized components are limited by their high cost and power consumption, while on-chip optical phased arrays face challenges in scaling and programmability. Here, we propose a solid-state approach to beam-steering using optomechanical antennas. We combine  recent progress in simultaneous control of optical and mechanical waves with remarkable advances in on-chip optical phased arrays to enable low-power and full two-dimensional beam-steering of monochromatic light. We present a design of a silicon photonic system made of photonic-phononic waveguides that achieves 44$^{\circ}$ field of view with $880$ resolvable spots  by sweeping the mechanical wavelength with about a milliwatt of mechanical power. Using mechanical waves as nonreciprocal, active gratings allows us to quickly reconfigure the beam direction, beam shape, and the number of beams. It also enables us to distinguish between light that we send and receive.

\end{abstract}

\pacs{Valid PACS appear here}
\maketitle
\footnote[]{$^{\dagger}$These authors contributed equally to this work.}
\footnote[]{$^{\star}$ sicamor@stanford.edu \\ rvanlaer@stanford.edu \\ safavi@stanford.edu}

\newpage

Fiber-coupled photonic circuits are powerful tools in our information infrastructure. In order to leverage these circuits to form and analyze light in our environment, we need to control how they radiate and absorb radiation. Gratings are an established way of controlling how photonic circuits radiate. They are dispersive: tuning the wavelength of light incident on a grating changes the angle at which it is scattered. Gratings can thus steer light in one dimension with a tunable laser~\cite{VanAcoleyen2011,Doylend2012}. When incorporated with phase-shifters into an array, they can steer in two dimensions \cite{Doylend2011,VanAcoleyen2011b,Hulme2015,Poulton2016,Sun2013,Heck2016}. These integrated beam-steering systems are of a size, weight, and cost surpassing the motorized optical gimbals currently used with free-space optical systems for lidar, optical wireless communication~\cite{Kedar2004,Elgala2011}, and free-space optical interconnects \cite{Rabinovich2015}. The growing presence of autonomous systems, such as self-driving cars, motivates the development of mass-manufacturable photonic systems. With low-power on-chip beam-steering, a host of remote sensing, communication, and display applications comes into reach.

\begin{figure}[ht]
\includegraphics[width=\linewidth]{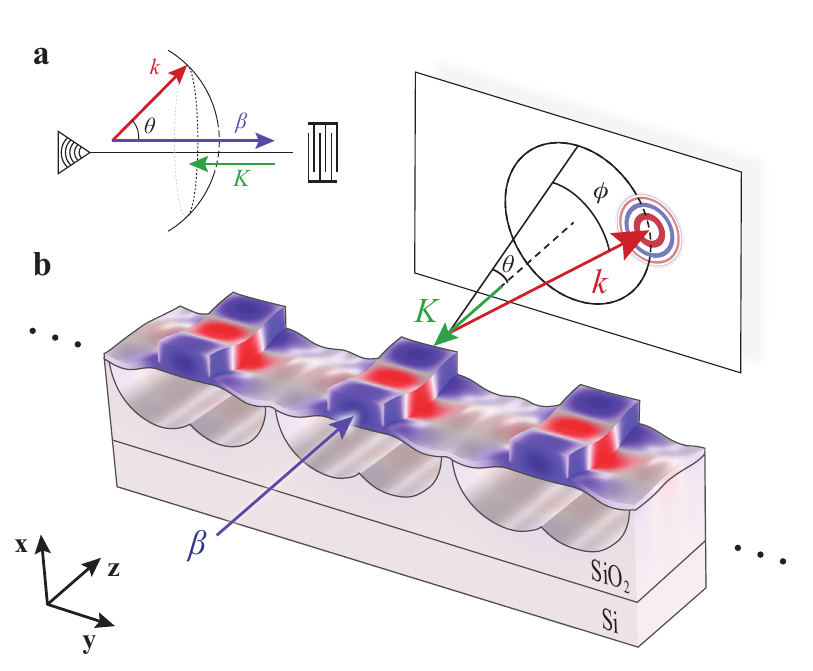}
\caption{\textbf{Optomechanical antenna arrays.} \textbf{a}, A mechanical wave with wavevector $K$ scatters a guided optical wave ($\beta$) into free-space ($k$) at an angle $\theta$.   Sweeping the mechanical frequency, and therefore $K$, steers a beam through a range of angles $\theta$. \textbf{b}, Incorporating these antennas into a phased array forms a beam in the far-field directed into angle $\phi$.  Antennas consist of partially released silicon ridge waveguides each of which supports guided optical and guided mechanical modes.  The displacement field $\vecb{u}$ that scatters guided light (color representing the electric displacement field $D_x$) out at $\theta = 60^\circ$.}
\label{fig:schem}
\end{figure}

With angular dispersion the etched grating's period \(\Lambda\) fixes a relation between incident wavelength \(\lambda\) and scattering angle \(\theta\). In contrast, with the ability to tune the grating period \(\Lambda\), a monochromatic beam can be formed and directed. Sound is a naturally tunable optical grating. An acoustic wave containing multiple wavelengths scatters light into multiple angles as realized in pioneering work on acousto-optic beam deflectors~\cite{Quate1965,Gordon1966,Korpel1966}. A progression to guided-wave, collinear systems ~\cite{Gfeller1977,Matteo2000} and arrays~\cite{Smalley2013} in Ti-diffused and proton-exchanged lithium niobate waveguides enabled large fields of view for monochromatic light. These low index-contrast lithium niobate waveguides limit integrability, resolution, and efficiency. We address these limitations by embracing high index-contrast, subwavelength-scale silicon waveguides to be incorporated into a dense phased array.  These waveguides -- engineered to guide both light and sound -- have recently been shown to exhibit strong acousto-optic interactions between propagating waves with tailorable dispersion~\cite{Kittlaus2015,VanLaer2015,Sarabalis2016}.

Here we develop the concept of an \emph{optomechanical antenna} (OMA) and present a perturbative description of the coupling between guided and radiated light by sound analogous to cavity optomechanics and Brillouin scattering \cite{Aspelmeyer2014,Eggleton2013,VanLaer2015,VanLaer2015b,Kittlaus2015,Sarabalis2016}. After illustrating the scattering process for a slab waveguide, we explore the optical and mechanical co-design of an OMA compatible with silicon photonics and practical for a phased array antenna. Such an OMA can scatter light in millimeters with only hundreds of microwatts of mechanical power. We conclude with an outline of this device's performance, an account on the effect of disorder, and an outlook on the new capabilities of this approach.

\section{Mechanics as a dynamic grating}
\label{sec:singleOMAntenna}

In an optomechanical antenna (Fig.~\ref{fig:schem}a), a guided optical wave with electric field  \(\vecb{E}\,\exp\prens{i\beta z - i\omega t}\) is scattered by a guided mechanical wave with displacement field  \(\vecb{u}\,\exp\prens{-iKz - i\Omega t}\) into radiating light with electric field $\vecb{E}_\trm{r}\,\exp\prens{i\vecb{k}\cdot\vecb{r} - i\omega_\trm{r}t}$. 
Energy and momentum conservation for the counter-propagating, anti-Stokes process 
\begin{align}
\label{eq:phasematching}
\omega_\trm{r} &= \omega + \Omega\\
k\,\cos\theta &= \beta - K,
\end{align}
determine the scattering angle $\theta$ (Fig.~\ref{fig:schem}a). The equations above describe the copropagating, Stokes process by reversing the sign of $\Omega$.
Under these phase-matching constraints, a single antenna radiates into a cone and a phased array into a pair of beams above and below the array. A frequency-swept mechanical drive sweeps the beam angle $\theta$ across the field of view in microseconds -- the time it takes a mechanical wave to traverse the antenna. 

Analogous to the treatment of interactions in cavity optomechanics and Brillouin scattering, we perturbatively compute radiation from an OMA.  Mechanical deformations vary the dielectric permittivity $\perm \rightarrow \perm + \delta_u\perm\cdot\vecb{u}$ with photoelastic and moving-boundary contributions to the scattering (see SI). Making a first Born approximation (see SI)~\cite{Saleh2007}, we have
\begin{equation}
\prens{\nabla\times\nabla\times - \mu\perm\omega^2} \E_\trm{r} = i\omega\mu \vecb{J}_\trm{om}
\label{eq:Ax=b}
\end{equation}
where the optomechanically-induced polarization current
\begin{equation}
\vecb{J}_\trm{om} = -i\omega\prens{\delta_u\perm\cdot\vecb{u}} \E
\end{equation}
is defined in terms of the unperturbed guided optical and mechanical modes $\E$ and $\vecb{u}$. The result is a set of inhomogeneous equations which we solve for the radiated electric field $\E_{\text{r}}$.  

Coupling to radiation causes decay of the optical power $\mathcal{P}(z)$ in the waveguide. 
From perturbation theory we find that the coupling between $\E$ and $\E_\trm{r}$ scales as $u=\text{max} |\m u|$. 
By Fermi's Golden Rule, the radiated optical power per unit length is $\alpha =\alpham u^2$ and therefore proportional to power in the mechanical wave $\mathcal{P}_\trm{m}$. Neglecting mechanical decay,   
\begin{equation}
	\mathcal{P}\prens{z} = \mathcal{P}_0 e^{-\alpham u^2 z}.
    \label{eq:opticalPowerAttenuation}
\end{equation}
The scattering rate $\alpham$ is the rate of conversion between guided and radiating optical fields per unit length per the square of the mechanical deformation amplitude. 
In contrast to static gratings where the scattering rate is fixed by fabrication, modulating the mechanical power modulates the effective aperture $L_\trm{eff} = \alpham^{-1} u^{-2}$ of an OMA.

\begin{figure*}
    \includegraphics[width=1.0\textwidth]{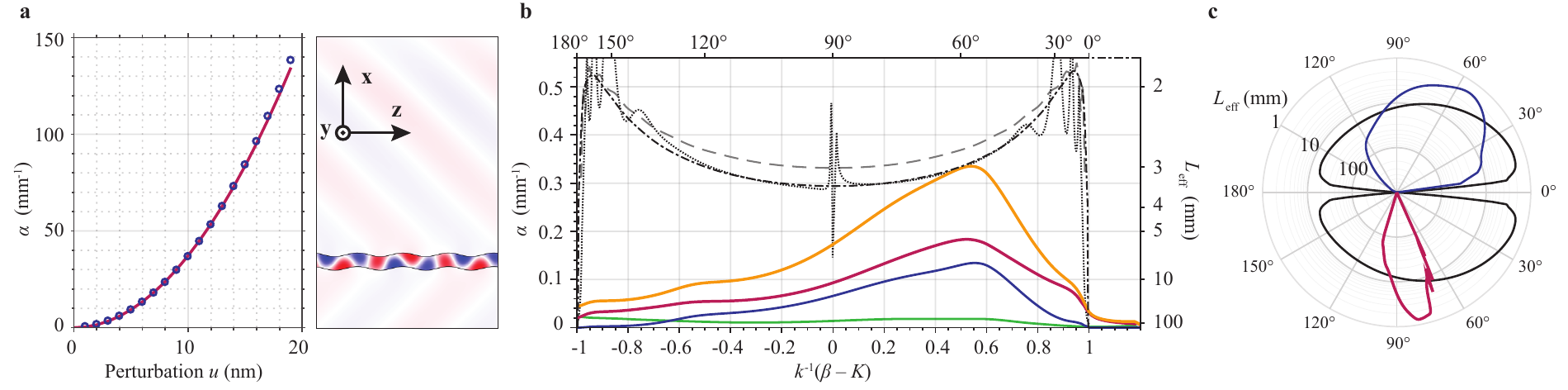}
\caption{\textbf{Optomechanical scattering.}
\textbf{a}, Scattering from the TE-polarized optical mode of a 220~nm thick suspended silicon slab is computed nonperturbatively and the electric displacement field $D_y$ is plotted where the maximal displacement $u$ is set to 30 nm for visibility of the radiated field. The single-sided scattering rate $\alpha$ is quadratic in $u$ even for such large displacements.  
\textbf{b}, The displacement-normalized scattering rates for the slab and ridge OMA are plotted together for $u = 1~\nm$. Slab rates plotted in black and gray are computed nonperturbatively in 2D (dotted) as well as perturbatively in 2D (dot-dash) and 3D (dashed).  For the ridge OMA of section~\ref{sec:OMAsforsiliconphotonics}, radiation into the optical slab modes (green), air (blue), and the silicon handle (red) add to give the total scattering rate (yellow). 
\textbf{c}, Normalizing the scattering rate by power, here 1~mW per antenna for a ridge OMA array and 1~mW per $1.41~\um$ for the slab, gives a practical measure of the field of view $\Delta\theta$. Tight confinement of the optical and mechanical energy in the transverse direction for the ridge OMA leads to enhanced radiation for a range of angles compared to the slab case.}
 \label{fig:OM_scattering}
\end{figure*}

\section{Radiation of a slab waveguide}
\label{sec:2domscattering}

We begin by analyzing a simple OMA: a 220~nm thick silicon slab waveguide suspended in air. 
A typical scattering process is plotted in Fig.~\ref{fig:OM_scattering}a where an antisymmetric mechanical Lamb wave scatters a counter-propagating guided transverse-electric (TE) optical mode into free space at $\theta = 45^\circ$.

The slab waveguide is simple enough to admit to an analytical approach.
A full coupled-mode description for roughness-induced scattering has been developed and is applicable to optomechanical scattering~\cite{Marcuse1969,Marcuse1969b}.
We take a numerical approach easily extended to arbitrary geometries in which finite-element-method solutions of the uncoupled equations drive the inhomogeneous equation~(\ref{eq:Ax=b}).

The scattering rates plotted in Fig.~\ref{fig:OM_scattering}b show that nanometer-scale oscillations yield millimeter-scale effective apertures. 
 Light which propagates with an effective index $n_\trm{eff} = 2.8$ scatters out symmetrically above and below the waveguide in the phase-matched region when $K/2\pi$ is between 1.2 $\mu\text{m}^{-1}$ ($\theta = 0^\circ$) and 2.5 $\mu\text{m}^{-1}$ ($\theta = 180^\circ$).
The moving-boundary term dominates the optomechanical interaction while the photoelastic contribution is $50\times$ smaller for this OMA (see SI). The interaction between TE guided light and Lamb waves is captured by an optomechanically-induced polarization current along $\hat{y}$
\[\vecb{J}_{\trm{om}} = -i\omega\Delta\perm E_y u_x \hat{y},\]
which drives $\hat{y}$-polarized TE optical fields in the surrounding medium.
For antisymmetric Lamb waves, the polarization currents induced on the top and bottom surfaces of the waveguide are $180^\circ$ out-of-phase, but since $n  t_\trm{Si} \approx \frac{\lambda}{2}$, they interfere constructively giving rise to strong scattering rates $\alpha$.
For the same reason, surface currents of symmetric Lamb waves interfere destructively such that the moving-boundary contribution to $\alpha$ is small (see SI).

Since the mechanical frequencies are much smaller than the optical frequency, they can be treated quasi-statically. Rather than the perturbative approach, the waveguide can be statically deformed by $\vecb{u}$ and solved for the radiating field by frequency or time-domain methods. The quasi-static, nonperturbative approach agrees well with perturbative calculations and results from literature (see SI).

The displacement-normalized $\alpham$ hides an important aspect of antenna performance: the mechanical power. 
The fixed-power antenna functions (Fig.~\ref{fig:OM_scattering}c) fall rapidly at higher angles $\theta$ since $K$ and therefore  $\mathcal{P}_\trm{m}\propto \Omega^{2}$ increase.
In comparison to OMAs in lithium niobate which employ surface acoustic waves, suspended structures are compliant and tightly confine the mechanical energy of their modes enabling orders-of-magnitude lower mechanical powers $\mathcal{P}_\trm{m}$.

\section{Optomechanical antennas for silicon photonics}
\label{sec:OMAsforsiliconphotonics}

\begin{figure}[htbp]
    \centering
    \includegraphics[width=1.0\linewidth]{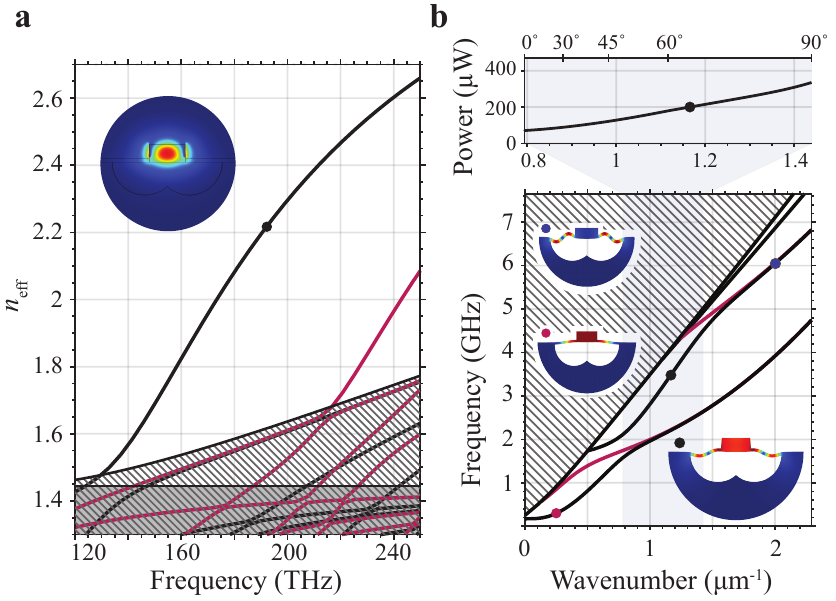}
    \caption{\textbf{Optics and mechanics of a ridge OMA.} The ridge OMA guides light and sound supporting a range of mechanical modes over which the 193 THz optical mode (dot on black band of \textbf{a}, \(E_y\) inset) can be scattered from \(0^\circ\) to \(90^\circ\) (shaded in blue). The mechanical power needed to achieve \(\max\left|u_z\right| = 1~\nm\) oscillations of the core is plotted above the bands of \textbf{b}, Inset plots of the displacement show the low \(K\) behavior (red) where the mode resembles the fundamental Lamb mode of a uniform, clamped membrane. The first excited symmetric mode in the shaded region (black dot) is the mode of interest with scattering rates plotted in Fig.~\ref{fig:OM_scattering}. After the anticrossing the mode is expelled from the core into the sockets (blue dot).  In \textbf{a}, the antisymmetric (with respect to $y$-reflection) and symmetric bands are plotted in black and red, respectively.  In \textbf{b}, symmetric and antisymmetric bands are plotted in black and red, respectively. In both band diagrams, the continuum of radiation modes of the 50~nm SOI stack are hatched.}

    \label{fig:codesign}
\end{figure}

Having explored a two-dimensional optomechanical antenna, we add transverse structure to our calculations to yield a design for an OMA practical for an array and compatible with microelectronics  manufacturing. 
In doing so, the mechanical waves of the slab are replaced by a multi-moded mechanical response of the core and socket of the waveguide in Fig.~\ref{fig:schem}b. Mixing between core and socket modes becomes an important feature of the antenna's optomechanical response.

The OMA we describe is designed for 220~nm silicon-on-insulator (SOI) common in silicon photonics.  A ridge waveguide is defined by a 170~nm partial etch, leaving a compliant 50~nm socket connecting the 450~nm wide optical core to the substrate.
The ridge is partially released leaving a $1.33~\um$ wide suspended region.
This width allows for subwavelength $1.41~\um$ pitched arrays with a transverse field of view $\Delta\phi$ of up to $67^\circ$.

This OMA supports both guided optical and guided mechanical modes.
At 193~THz optical waves are confined to the waveguide core and have an effective index $n_\trm{eff} = 2.33$, well above the slab modes of the 50~nm SOI stack (hatched in Fig.~\ref{fig:codesign}a). 
The effective index sets the range of wavevectors $K \in 2\pi\times\left\lbrack 0.86, 2.15 \right\rbrack \um^{-1}$ that phase-match $\beta$ to free-space $k$. For these wavevectors, Lamb-like flexural modes of the waveguide have lower phase velocities than any other mechanical excitation in the system. They comprise bands that fall below the cone of surface and bulk waves of the $50 \, \text{nm}$ SOI stack, represented by the hatched region in Fig.~\ref{fig:codesign}. Consequently, they do not suffer mechanical radiation losses in the absence of disorder~\cite{Sarabalis2016,Sarabalis2017}.

Mechanical modes of the structure can be understood in terms of waves in the sockets and waves in the core. Only motion of the core causes the antenna to radiate, directing our attention to the first excited symmetric band of Fig.~\ref{fig:codesign}b. The fast band of the core mixes with the slow bands of the socket giving rise to avoided crossings at $K/2\pi$ of $0.8~\um^{-1}$ and $1.5~\um^{-1}$. Below and above the avoided crossings, the sockets are mechanically decoupled by the core. This leads to nearly degenerate symmetric (black) and antisymmetric (red) bands. The motion of the core is suppressed in these degenerate bands, and therefore there is no optomechanical coupling. 

The scattering rates are computed perturbatively and plotted in Fig.~\ref{fig:OM_scattering}b alongside the results for a slab. The scattering rate $\alpha$ is peaked at $\theta \approx 60^\circ$  and falls off near the avoided crossings. Since the scattering rates are normalized by the maximum displacement of the core and not by power, these tails come from changes to the mechanical mode profile. Radiation into air is slightly weaker than into silicon. The latter is small but nonzero even when radiation into air is disallowed by phase-matching $\left|k^{-1}\prens{\beta - K}\right| > 1$. The ridge OMA scatters out at nearly half the rate of the slab.

Despite the effects of mechanical mode mixing, the ridge OMA retains a large field of view as shown by the power-normalized antenna function of Fig.~\ref{fig:OM_scattering}c. For less than a milliwatt of mechanical power per antenna, corresponding to approximately $1~\text{nm}$ maximal displacement, scattering lengths of $2~\trm{mm}$ can be achieved over $\Delta\theta = 13^\circ$.  Doubling either the power or the scattering length more than doubles the field of view $\Delta\theta = 44^\circ$.

\section{Optics of optomechanical antennas}
\label{sec:opticsOMA}

\begin{table}[t]
\centering
\setlength{\tabcolsep}{12pt}
\renewcommand{\arraystretch}{2.0}
\begin{tabular}{c | c}
 Antenna property &  \\ \hline
Antenna length $L_{\text{eff}}$ 		& 2 mm \\
Mechanical power $\mathcal{P}_\trm{m}$ & 2 mW \\
Transducer bandwidth $\frac{\Delta\Omega}{2\pi}$		& 1.6~GHz \\
Field of view $\Delta\theta$		& $44^\circ$  \\
Spot size $\delta\theta$		& $0.05^\circ$   \\
Resolvable spots $N_{\theta}$		& $880$    \\
Optical bandwidth $\frac{\Delta\omega}{2\pi}$		& 39~GHz   \\
Mechanical bandwidth $\frac{\Delta\Omega_{\text{m}}}{2\pi}$		&   2~MHz
\end{tabular}
\caption[Optomechanical antenna properties.]{\textbf{Optomechanical antenna properties.} We assume an effective antenna length limited by the expected dephasing lengths described in the text. The field of view is from $\theta$ of $35^{\circ}$ to $79^{\circ}$. We compute the optical bandwidth around $\theta = 60^{\circ}$ with an optical group index of $4.3$ and a mechanical group velocity of $4135 \, \text{m/s}$.}
\label{table:figuresofmerit}
\end{table}

\begin{table*}[t]
\centering
\setlength{\tabcolsep}{3pt}
\renewcommand{\arraystretch}{1.8}
\begin{tabular}{c | c c c | c c c}
 Geometric dephasing	 	&  $\sigma_{l} \, $(nm) & $\partial_l \beta \, \, ((\text{mm nm})^{-1})$ & $\partial_l K \, ((\text{mm nm})^{-1})$ & $L_{\varphi,\beta} \, $(mm) & $L_{\varphi} \, $(mm) & $L_{\varphi,\text{co}} \, $(mm) \\ \hline
Core width & 0.5 & 7.8 & -1.2 & 13 & 13 & 9.7\\
Core height & 0.3 & 14.6 & -12.2 & 10 &  6.1 & 3.1\\
Slab height & 0.5 & 9.8 & -11.9 & 8.1 & 3.3 & 1.7 \\
Membrane width & 2 & $-0.003$ & 1.2 & 4021 & 35 & 34 \\
\hline\hline
Thermal dephasing	 	&  $\Delta T \, $(K) & $\partial_T \beta \, \, (\text{(mm K)}^{-1})$ & $\partial_T K \, (\text{(mm K)}^{-1})$ & $L_{T,\beta} \, $(mm) & $L_{T} \, $(mm) & $L_{T,\text{co}} \, $(mm)\\ \hline
 & 5 & 0.9 & 0.2 & 1.4 & 1.1 & 1.8 \\

\end{tabular}
\caption[Expected dephasing effects.]{\textbf{Optical and mechanical phase errors set a millimeter-scale upper bound on the length of the antenna.} We compute the optical, mechanical and total dephasing induced by nanoscale disorder (top rows) and by thermal gradients (bottom row) across the antenna array. The standard deviations $\sigma_l$ are based on measurements of similar silicon photonic circuits \cite{Elson1995,Selvaraja2009,Fursenko2012,Melati2014}. We compute the mechanical sensitivities at a wavelength of $\Lambda = 842 \, \text{nm}$ and frequency of $\Omega/2\pi = 3.42 \, \text{GHz}$ such that the steering angle $\theta = 60^{\circ}$ corresponds to the maximal scattering rate $\alpha$ (Fig. \ref{fig:OM_scattering}). The dephasing lengths are computed with equations given in the main text, estimating the correlation lengths $\xi_l$ at $50 \, \mu\text{m}$ (see SI). The mechanical and optical sensitivities are usually of opposite sign: mechanical waves speed up when optical waves slow down. Both are most vulnerable to height disorder and with comparable sensitivities. In the purely optical case, the overall geometric dephasing length is $L_{\varphi,\beta} = 3.4 \, \text{mm}$. For counter-propagating optical and mechanical waves we have $L_{\varphi} = 1.7 \, \text{mm}$: about half of $L_{\varphi,\beta}$ as the optical and mechanical phase errors add incoherently. Further, the dephasing length for copropagating fields $L_{\varphi,\text{co}} = 0.9 \, \text{mm}$ is about four times smaller than $L_{\varphi,\beta}$ as the optical and mechanical phase errors subtract coherently. Finally, both mechanical and optical fields slow down with increasing temperature in silicon, generating millimeter-scale dephasing lengths in either case.}
\label{table:dephasing}
\end{table*}

In the previous sections we designed an optomechanical antenna that can be integrated into a silicon photonic phased array. Here we discuss the main properties of its radiation pattern.  

In the far-field the beam radiated from an OMA array is governed by Fraunhofer diffraction. For an ideal radiator where $\vecb{J}_\trm{om}$ does not vary in the longitudinal direction and remains coherent over a length $L_\trm{eff}$, the far-field spot size is $\delta\theta = \lambda/( L_\trm{eff}\sin\theta)$. The polar field of view $\Delta\cos\theta = \Delta K/k$ is set by the range of wavevectors $\Delta K$ for which $\alpha$ is large as quantified in Fig.~\ref{fig:OM_scattering}c. A full system requires an efficient mechanical transducer and coupling structure with bandwidth $\Delta \Omega$ over this range. Assuming negligible mechanical group velocity dispersion, the number of resolvable spots is the mechanical transit time-bandwidth product $N_\theta = \Delta \theta/\delta \theta = \tau_\trm{m}\Delta\Omega/2\pi$. Estimates for the ridge OMA are in table \ref{table:figuresofmerit}.
Since $L_\trm{eff}$ plays an important role in beam quality, we quantify sources of spatial decay and decoherence of $\vecb{J}_\trm{om}$ that limit $L_\trm{eff}$. Variations in the amplitude and phase of $\vecb{J}_\trm{om}$ arise from optical and mechanical decay, as well as dephasing due to geometric disorder and thermal fluctuations.  

As light and sound propagate along the antennas, the phase of $\vecb{J}_\trm{om}$ accumulates an error $\delta \varphi(z)$.  This differs from beam-steering systems that use spatial light modulators \cite{Engstrom2008}, MEMS micromirrors \cite{Yoo2013,Megens2014}, or microlens arrays \cite{Tuantranont2001} where light interacts with the device only over a small distance.
For fluctuations $\delta \beta(z)$ and $\delta K(z)$ spatially correlated over $\xi \ll L_\trm{eff}$, the phase error $\delta\varphi(z)$ diffuses along the antennas and is Gaussian-distributed with its variance growing linearly with $z$.  
We define $L_{\varphi}$ as the length after which the phase variance $\langle \delta \varphi^{2}(z) \rangle$ averaged along the antennas equals $\pi^{2}$.  
This dephasing length $L_{\varphi}$ depends on the relative propagation direction of the guided optical and mechanical waves. In the counter-propagating case we find
\begin{equation}
\left\langle\delta\varphi^2\prens{z}\right\rangle = S_{\beta\beta}[0] z + S_{KK}[0](L-z)
\end{equation}
where $S_{\beta\beta}$ and $S_{KK}$ are the power spectral densities of $\delta\beta$ and $\delta K$ (see SI).

Geometric fluctuations $\delta X_l$, indexed by $l$, shift $\beta$ by $\delta \beta(z) = \sum_l \partial_l \beta \delta X_l(z)$ such that for stationary noise with correlations
\begin{equation}
	\left\langle \delta X_l\prens{\Delta z}\delta X_l\prens{0}\right\rangle = \sigma_l^2e^{-\left|\Delta z\right|/\xi_l}
	\label{eq:noiseCorrelations}
\end{equation}
we get $S_{\beta\beta}[0] = 2\sum_l (\partial_l \beta \sigma_l)^{2} \xi_l$ and similarly for $K$. Slow drifts in $X_l$ (large $\xi_l$) are more limiting than roughness since they lead to more rapid phase accumulation along each antenna.  Therefore the dephasing length is
\begin{equation}
	L_{\varphi} = \frac{2\pi^2}{S_{\beta\beta}[0] + S_{KK}[0]}
	\label{eq:Leffphi_counterAntiStokes}
\end{equation}
In the copropagating case we similarly obtain $
\langle \delta \varphi^{2}(z)\rangle = S_{k_{||}k_{||}}[0]z$ with  $\delta k_{\shortparallel}(z) = \delta\beta(z) - \delta K(z)$, and the dephasing length $L_{\varphi,\trm{co}}$ is found by replacing the denominator of Eqn. (\ref{eq:Leffphi_counterAntiStokes}) by $S_{k_{||}k_{||}}[0]$.

We compute the geometric sensitivities to fluctuations $\partial_l \beta$ and $\partial_l K$ for different types of pertubations and find height variations to be the dominant source of dephasing.  Our finite-element models predict similar optical and mechanical sensitivities to height disorder (table \ref{table:dephasing}). 
A counter-propagating optomechanical antenna array dephases after $L_{\varphi} = 1.7 \, \text{mm}$, approximately half a purely optical array where $L_{\varphi} = 3.4 \, \text{mm}$.

Temperature gradients across the system also shift the phases of the guided optical and mechanical fields. Spatially inhomogeneous temperatures are analogous to geometric disorder. A temperature gradient $\Delta T$ across the antenna array results in thermal dephasing lengths of $L_{\text{T}} = 2\pi/(|\partial_T(\beta + K)| \Delta T)$ in the counterpropagating and $L_{T,\text{co}} = 2\pi/(|\partial_T k_{\shortparallel}| \Delta T)$ in the copropagating case (see SI). Careful thermal management can likely limit spatial temperature gradients to $5 \, \text{K}$ across the array \cite{Zhang2014} so we find $L_{\text{T}} \approx 1.5 \, \text{mm}$ in either case (table \ref{table:dephasing}).

Optical and mechanical crosstalk between the waveguides in a phased array is another source of phase errors. Crosstalk splits the wavevectors of symmetric $k_+$ ($\dots + + + + \dots$) and antisymmetric $k_-$ ($\dots + - + - \dots$) array supermodes, causing them to dephase after a length $L_\text{x} = \pi/(k_+ - k_-)$. A $1.41~\um$ pitch array has an optical crosstalk length of $L_\text{x} = 8.7~\mm$.

Larger apertures yield narrower spots and higher resolution at the cost of the optical and mechanical modulation bandwidth. Although our control over mechanical wavevector $K$ enables beam-steering at fixed optical frequency, optomechanical antennas are still dispersive. The optical bandwidth $\Delta \omega$ at fixed mechanical frequency is set by how much $\omega$ can be changed before a spot shifts by $\delta \theta$.
We find $\Delta\omega/2\pi = 1/(\tau - \tau_{\text{r}})$ with $\tau = L_{\text{eff}} n_{\text{g}}/c$ the transit time of the guided optical wave where $n_\text{g}$ is the group index, and $\tau_{\text{r}} = L_{\text{eff}} \cos{\theta}/c$ the transit time of the radiating field across the aperture (see SI). Similarly, the mechanical transit time $\tau_{\text{m}}$ determines the mechanical bandwidth $\Delta \Omega_{\text{m}}/2\pi = 1/\tau_{\text{m}}$ within a spot at fixed optical frequency. We provide estimates for these antenna properties in table~\ref{table:figuresofmerit}. 

\begin{figure}[ht]
\includegraphics[width=\linewidth]{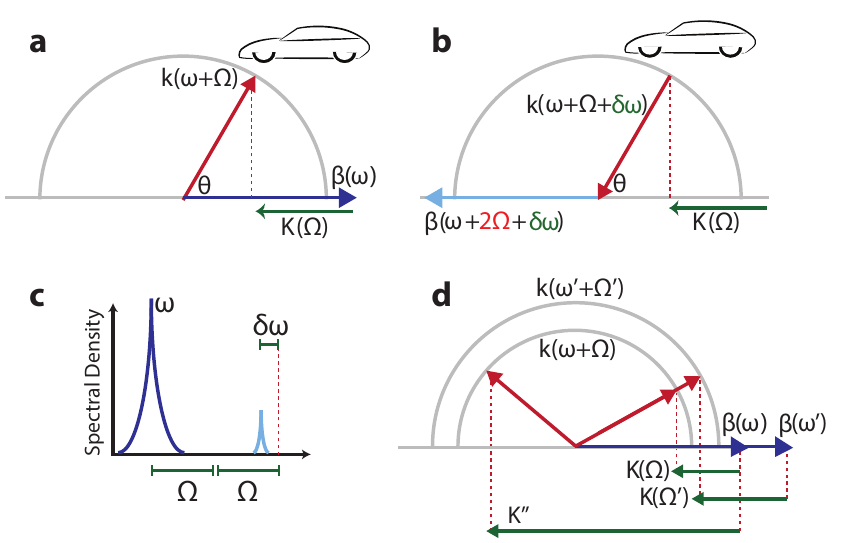}
\caption{\textbf{Automatic frequency duplexing of transmit and receive and dynamical multiplexing of outgoing beams.} \textbf{a}, On the transmit side, the optical wave at $\omega$ is shifted to an outgoing optical wave at frequency $\omega+\Omega$. \textbf{b}, The backscattered light now has an additional Doppler shift of $\delta\omega$. This incoming wave mixes with the same mechanical wave and excites the optical waveguide at frequency $\omega+2\Omega+\delta\omega$. \textbf{c}, The resulting optical spectrum allows us to distinguish between light being sent and received, moving the received beam outside of the laser phase noise. \textbf{d}, The outgoing radiation can be multiplexed dynamically by exciting a superposition of mechanical waves. Each of the outgoing beams can be controlled independently. As a special case, this implies widely different optical wavelengths can be sent to and received from the same spot.}
\label{fig:outlook}
\end{figure}

\section{Outlook}
\label{sec:outlook}
In addition to two-dimensional beam-steering, active gratings generated by mechanical waves naturally implement certain functionalities that are not intrinsically present in other approaches.

First, the time-varying grating generated by a unidirectional mechanical field breaks reciprocity in the structure. Therefore the system operates as a nonreciprocal metasurface \cite{Shi2017}: light that is scattered out is shifted up in frequency by $\Omega$ and the light coming back is shifted up \textit{again} in frequency by $\Omega$ such that the roundtrip optical frequency shift is $2\Omega$ (Fig. \ref{fig:outlook}). Therefore the system naturally includes a frequency-shifting function that can be used for heterodyning with an on-chip local oscillator.

Second, even at a fixed optical wavelength we can inject a superposition of mechanical waves with different wavevectors. Each of these mechanical waves generates an outgoing beam at a separate angle that can be controlled, sent to different targets and read out independently (Fig. \ref{fig:outlook}d). As a special case, this enables multiple optical wavelengths to be sent to and received from a single angular spot. Such functionality may prove useful in the realization of free-space communication links \cite{Rabinovich2015,Rabinovich2015a}, (holographic) video displays \cite{Smalley2013a,Korpel1966}, remote sensing \cite{Wang2007,Schliesser2012,Boudreau2013}, and coherent imaging \cite{Aflatouni2015,Fatemi2017}.
\\

\section{Conclusion}
\label{sec:conclusion}
In conclusion, we propose an on-chip, two-dimensional beam-steering system compatible with standard microelectronics processes based on guided mechanical waves. We design hybrid photonic-phononic waveguides whose mechanical excitations can travel on the surface of a silicon-on-insulator chip. The propagating mechanical fields -- acting as active gratings -- convert between guided and radiating optical fields in a rapidly reconfigurable way. Efficient optical mode conversion can be realized in millimeter-scale apertures with low mechanical drive power. 
The system can steer monochromatic light over a large field of view; distinguish between outgoing and incoming light through a nonreciprocal frequency shift; and control the beam direction, beam shape, and the number of beams. More generally, we have shown that subwavelength control of photons and phonons enables low-power, dynamic control of light. 

%

\vspace{5mm}
\textbf{Acknowledgement.} We acknowledge support by the National Science Foundation (ECCS-1509107), the Stanford Terman Fellowship and the Hellman fellowship, support from ONR QOMAND MURI, as well as start-up funds from the Stanford University school of Humanities and Sciences. R.V.L. acknowledges funding from VOCATIO and from the European Union's Horizon 2020 research and innovation program under Marie Sk\l{}odowska-Curie grant agreement No. 665501 with the research foundation Flanders (FWO). We thank Jeremy Witmer, Okan Atalar, Rishi Patel, and Patricio Arrangoiz-Arriola for discussions.

\textbf{Contributions.} All authors contributed to developing the concept, methods of analysis, and writing of the manuscript.

\appendix

\section{Simulation methods}
\label{sec:scatteringRate}

\subsection{Nonperturbative method of computing scattering rates}
\label{subsec:themodel}

We perform nonperturbative scattering simulations with the finite-element solver COMSOL~\cite{COMSOL5} solving Maxwell's equations in either 2D or 3D in the frequency domain.  The resulting eigenvalue problem has nearly guided solutions. Light scattered out of the slab is absorbed by a perfectly matched layer, causing the eigenvalue to become complex $\omega\rightarrow \omega- i\frac{\kappa}{2}$.    In this section we show that the scattering rate $\alpha$ is related to $\kappa$ as
\begin{equation}
\alpha = \frac{\kappa}{v_{\text{g}}} = \frac{\omega}{Q v_{\text{g}}}
\end{equation}
where $v_{\text{g}} = c/n_{\text{g}}$ is the optical group velocity.   

Consider an optical waveguide with energy per unit length \( \mathcal{E}, \)
 power transmitted down the waveguide \( \mathcal{P}, \) and power scattered out of the waveguide per unit length \( \mathcal{P}_s. \)
 We would like to relate \( \mathcal{P}_s \)
 to the attenuation rate \( \alpha. \)
 Energy conservation yields
\[ \partial_t \mathcal{E} = -\partial_z \mathcal{P} - \mathcal{P}_s \]
which for steady state \(\partial_t \rightarrow 0\) becomes \( \partial_z \mathcal{P} = - \mathcal{P}_s. \) Since
\[ Q = \frac{\omega}{\kappa}~~~\textrm{and}~~~\kappa = \frac{\mathcal{P}_s}{\mathcal{E}}, \]
our statement of energy conservation becomes
\[ \partial_z \mathcal{P} = -\frac{\omega}{Q}\mathcal{E} \]
which can be reexpressed using \( v_\text{g} = \frac{\mathcal{P}}{\mathcal{E}} \) as 
\[ \partial_z\mathcal{P} = - \underbrace{\frac{\omega}{Q v_{\text{g}}}}_{\equiv\alpha}\mathcal{P}. \]
Therefore, the optical power decays exponentially at a rate \(\alpha\) that can be expressed in terms of the imaginary part of the eigenvalues of our numerical solutions.

\subsection{Verification of the model}
\label{subsec:verificationmodel}

\subsubsection{Literature grating simulation}
\label{subsubsec:literaturegrating}

\begin{figure}[ht]
\includegraphics[width=\linewidth,trim=0cm 0cm 0cm 0cm,clip]{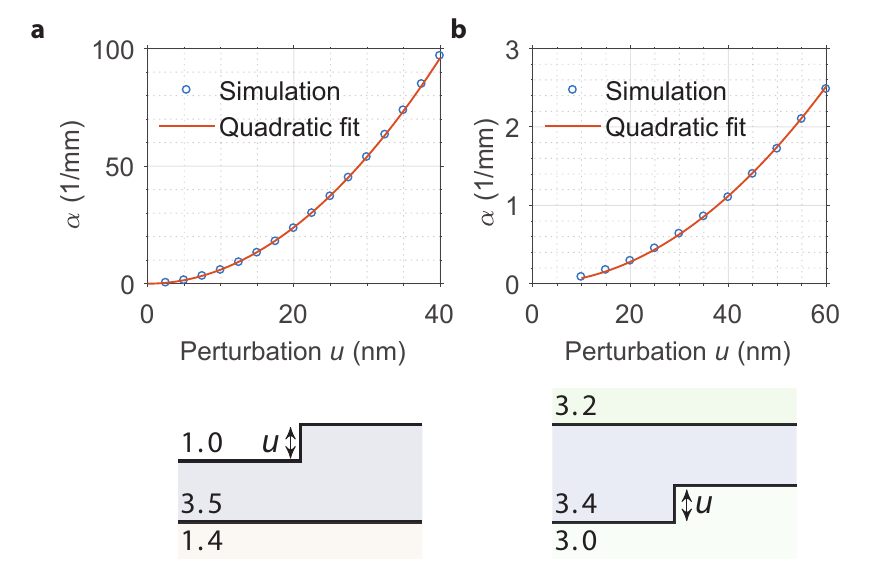}
\caption{\textbf{Simulation of literature grating couplers.} The scattering strength scales quadratically with the perturbation $u$. \textbf{a}, Simulation of strong SOI grating \cite{Taillaert2005} that has been tested experimentally \cite{Taillaert2005,Laere2007}. The group index of the optical mode is $n_{\text{g}} = 3.6$ and the Bloch index is 2.85. The grating pitch is $630 \, \text{nm}$ and the core thickness is $220 \, \text{nm}$. \textbf{b}, Simulation of weak grating \cite{Little1996}. The group index of the optical mode is $n_{\text{g}} = 3.4$ and the Bloch index is 3.29. The grating pitch is $10 \, \mu\text{m}$ and the core thickness is $500 \, \text{nm}$.}
\label{fig:verification}
\end{figure}

We test our nonperturbative computational approach by modeling two wavelength-scale periodic optical structures from the literature: (1) a high index-contrast, strong silicon-on-insulator grating coupler \cite{Taillaert2005} used in imec's silicon photonics pilot line \cite{Laere2007} and (2) a low index-contrast, weak grating coupler \cite{Little1996}.

\begin{table}[htbp]
\centering
\setlength{\tabcolsep}{4pt}
\renewcommand{\arraystretch}{1.8}
\begin{tabular}{c | c c}
	 			& $\alpham \, (\text{mm}^{-1} \text{nm}^{-2})$  & $ \alpham^{-1} (\text{mm} \, \text{nm}^{2})$ \\ \hline
SOI literature \cite{Taillaert2005,Laere2007} 		& 0.059  & 16.9  \\
Our model 		& 0.061 & 16.4  \\ \hline
Little \cite{Little1996} 		& $7.5 \cdot 10^{-4}$ & 1333  \\
Our model 		& $7.1 \cdot 10^{-4}$ & 1408  \\
\end{tabular}
\caption[Comparison of scattering rates to literature]{\textbf{Our model agrees with literature results.} References \cite{Taillaert2005} and \cite{Little1996} provide the scattering rate $\alpha$ as a function $u$ for these gratings. We compute $\alpham$ by estimating a couple of points in their figures in a similar range of $u$. Our model agrees with the literature result up to $3\%$ for the strong grating and up to $5.6\%$ for the weak grating.}
\label{table:comparisonliterature}
\end{table}

\subsubsection{PML tests}
\label{subsubsec:PMLtests}
\begin{figure}[ht]
\includegraphics[width=\linewidth,trim=0cm 0cm 0cm 0cm,clip]{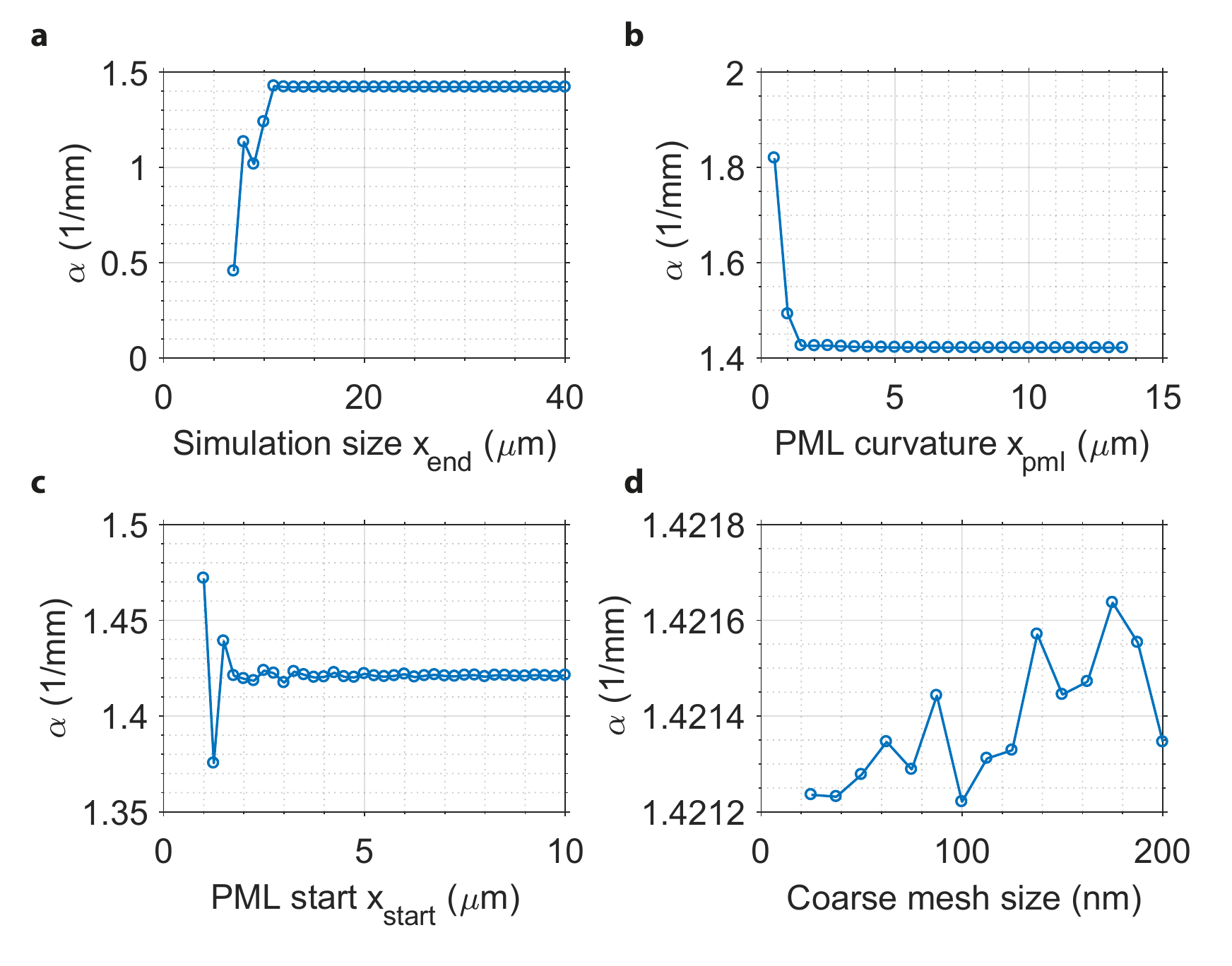}
\caption{\textbf{Study of the PML and simulation domain parameters.} \textbf{a}, Sweep of simulation size $x_{\text{end}}$ shows initially an increase and then saturation of the scattering rate $\alpha$ when $x_{\text{end}} > 10 \, \mu\text{m}$. \textbf{b}, The scattering rate drops fast with PML curvature $x_{\text{pml}}$ and saturates beyond $x_{\text{pml}} > 2 \, \mu\text{m}$. \textbf{c}, The scattering rate remains constant when $x_{\text{start}} > 3 \, \mu\text{m}$. \textbf{d}, Coarse mesh size does not strongly impact scattering rate below $200 \, \text{nm}$. Similarly, the fine mesh size does not affect the scattering rate below $20 \, \text{nm}$.}
\label{fig:pml_study}
\end{figure}

We implement the perfectly matched layers (PMLs) at the bottom and top of the simulation as an imaginary part of the refractive index $\Im{n}$ starting a distance $x_{\text{start}} = 6 \, \mu\text{m}$ above and below the waveguide. For $|x| > x_{\text{start}}$ the strength of the PML is set by
\begin{equation*}
\Im{n} = \left(\frac{|x| - x_{\text{start}}}{x_{\text{pml}}}\right)^{2}
\end{equation*}
such that $\Im{n} = 1$ at a distance $|x| = x_{\text{start}} + x_{\text{pml}} = 15 \, \mu\text{m}$ away from the waveguide. The computational domain ends at $|x| = x_{\text{end}} = \frac{d}{2} + 20 \, \mu\text{m}$ with perfectly conducting boundaries and $d = 220 \, \text{nm}$ the typical core thickness. We set the maximum mesh size in the core and free-space domains at $10 \, \text{nm}$ and $150 \, \text{nm}$. With these parameters, the simulation time is about $20 \, \text{s}$ for a wavelength of $0.63 \, \mu\text{m}$. We check PML operation with the following tests, all for the SOI grating coupler of Fig. \ref{fig:verification}a and for a perturbation of $u = 5 \, \text{nm}$. The scattering rate $\alpha \approx 1.42/\text{mm}$ in these tests (Fig. \ref{fig:verification}a and \ref{fig:pml_study}).

First, we investigate the field intensity in the free-space domain. For $1 \, \mu\text{m} < |x| <  x_{\text{start}}$ the field remains roughly constant, confirming that the PML mainly measures the radiation power and not the evanescent field of the guided mode. Next, we sweep the size of the computational domain $x_{\text{end}}$ (Fig. \ref{fig:pml_study}a). The scattering rate saturates fast, likely because of reduced reflections off the perfectly conducting boundaries, and stays constant up to $10^{-4}$ fractionally afterwards. Second, we sweep $x_{\text{pml}}$ and thus the PML strength (Fig. \ref{fig:pml_study}b). The scattering rate decreases rapidly until $x_{\text{pml}} > 2 \, \mu\text{m}$ and then fluctuates at $10^{-4}$ fractionally. Third, we sweep $x_{\text{start}}$ and $x_{\text{pml}} = x_{\text{start}} + 3 \, \mu\text{m}$ while keeping other parameters constant (Fig. \ref{fig:pml_study}c). The scattering rate oscillates initially and then stays constant up to $10^{-4}$ fractionally. Fourth, we sweep the maximum mesh element size in the free-space domains (Fig. \ref{fig:pml_study}c). The scattering rate again fluctuates at $10^{-4}$ level fractionally.

Our standard operating parameters are all chosen in these regions of $10^{-4}$ fractional sensitivity to discretization and PML parameters. Therefore, we expect simulation results accurate at percent-level at least for decay rates far above $\frac{\kappa}{2\pi} \approx 10 \, \text{MHz}$, quality factors below $Q \approx 10^{7}$, scattering rates above $\alpha \approx 10/\text{m}$ and decay lengths below $\alpha^{-1} \approx 10 \, \text{cm}$. We determine all scattering rates a factor $10^{3}$ to $10^{4}$ away from these thresholds. The quadratic scaling of scattering rate with respect to perturbations enables extrapolation to smaller perturbations $u$ where necessary.

\subsubsection{FDTD calculations}
\label{subsubsec:lumerical}
In addition to the COMSOL-based frequency-domain models, we also developed Lumerical-based finite-difference time-domain (FDTD)\cite{Lumerical} 2D and 3D models to compute the scattering rates and radiation patterns. Here we inject a pulse with a bandwidth of $5$ to $10 \, \text{nm}$ and determine the scattering rate $\alpha$ from the exponential decay of the guided power. Lumerical has built-in functions that allow for a straightforward determination of the electromagnetic beam in the far-field and thus the angles and strengths of the first-, second- and higher-order grating lobes. The results generally agree with the COMSOL-based frequency-domain approach described above and in the main text. We focus on the faster frequency-domain simulations.

\section{Silicon slab in air with sinusoidal perturbation}
\label{sec:siliconslabsinus}

\subsection{Scattering rate comparison}
\label{subsec:comparison}
In this section, we investigate the scattering rates of suspended silicon-on-insulator slabs in greater detail.

\begin{table}[htbp]
\centering
\setlength{\tabcolsep}{3pt}
\renewcommand{\arraystretch}{1.7}
\begin{tabular}{c | c c c}
	 			& $\alpha_{\text{m}} \, (\text{mm}^{-1} \text{nm}^{-2})$  & $ \alpham^{-1} (\text{mm} \, \text{nm}^{2})$ & $Q_{\text{m}} \, (\text{nm}^{2})$ \\ \hline
\raisebox{-0.4\totalheight}{\includegraphics[width=0.15\linewidth,trim=5.2cm 25cm 14.3cm 1.7cm,clip]{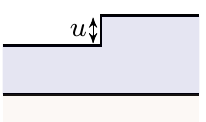}} 		& 0.059  & 16.9  & $2.5 \cdot 10^{5}$\\
\raisebox{-0.4\totalheight}{\includegraphics[width=0.15\linewidth,trim=5.2cm 25cm 14.3cm 1.7cm,clip]{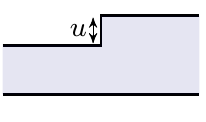}}  							& 0.073  & 13.7  & $2.0 \cdot 10^{5}$\\
\raisebox{-0.4\totalheight}{\includegraphics[width=0.15\linewidth,trim=5.2cm 24.7cm 14.3cm 1.7cm,clip]{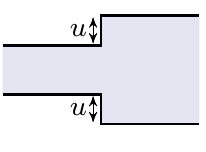}} 					& $1.4 \cdot 10^{-6} \left(u(\text{nm})\right)^2$  & $\frac{7.1 \cdot 10^{5}}{\left(u(\text{nm})\right)^2}$   & $\frac{10^{10}}{\left(u(\text{nm})\right)^2}$\\
\raisebox{-0.4\totalheight}{\includegraphics[width=0.15\linewidth,trim=5.2cm 24.7cm 14.3cm 1.7cm,clip]{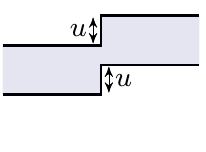}} 					& 0.31  & 3.2  & $5.8 \cdot 10^{4}$\\
\raisebox{-0.4\totalheight}{\includegraphics[width=0.15\linewidth,trim=5.2cm 24.7cm 14.3cm 1.7cm,clip]{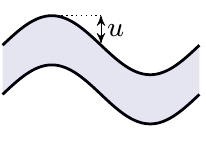}} 	& 0.74 & 1.3 & $2.4 \cdot 10^{4}$ \\
\end{tabular}
\caption[Scattering rate comparison of various perturbations]{\textbf{Scattering rate comparison of various perturbations.} The scattering rate varies several orders of magnitude depending on the type of perturbation. A symmetric, sinusoidal perturbation has the largest scattering rate in a $220 \, \text{nm}$ thick silicon slab. For some perturbations the scattering rate scales with $u^{4}$ instead of $u^{2}$ because of destructive cancellations in the term proportional $u^{2}$. In general the scattering rate contains all terms $u^{2l}$ with $l$ any positive integer. Here we use the perturbation-normalized quality factor as $Q_{\text{m}} = \omega/(\alpham v_{\text{g}})$.}
\label{table:comparison}
\end{table}

Table \ref{table:comparison} shows a comparison of the scattering rates of five types of perturbations to a $220 \, \text{nm}$ thick silicon slab waveguide. We perform these calculations at fixed frequency of $193.5 \, \text{THz}$ and fixed scattering angle of $67^{\circ}$. The Bloch index of the optical slab mode is about 2.82 and the grating pitch is $630 \, \text{nm}$. From top to bottom, the first grating is the silicon-on-insulator grating of section \ref{subsubsec:literaturegrating}. It has only a top surface rectangular perturbation. The second grating has air both above and below the core. Its scattering rate is slightly higher owing to the increased index-contrast. The third structure has a symmetric perturbation to the top and bottom surfaces. Its scattering rate for a $\delta = 1 \, \text{nm}$ perturbation is about four orders of magnitude below that of the SOI grating coupler that breaks vertical symmetry. In addition, its scattering rate scales with $\delta^{4}$ instead of the usual $\delta^{2}$ -- even though second-order scattering (satisfying $\beta - 2K = k\cos{\theta_{2\text{nd}}}$) is not phase-matched for any angle $\theta_{2\text{nd}}$ when $\theta = 67^{\circ}$. The suppressed scattering and $\delta^{4}$-scaling arise from destructive cancellations in the fields scattered from the top and bottom surfaces, see Fig. \ref{fig:thickness_sweep} and discussion below for details. The fourth grating is identical except in that it has an antisymmetric perturbation to the top and bottom surfaces. The scattering rate $\alpha$ is a factor $\frac{0.31}{0.073} = 4.3$ larger than that of a grating with a perturbation only on the top surface. The enhanced scattering arises from constructive interference between the fields radiated by top and bottom perturbations (Fig. \ref{fig:thickness_sweep}). The fifth grating has a sinusoidal instead of a rectangular perturbation. It is nearly identical to the dynamic mechanical field we propose to excite. The first Fourier component of a rectangular signal is $\frac{2}{\pi}$, so the scattering by the sinusoidal perturbation is a factor $\left(\pi/2\right)^{2} = 2.4$ stronger. The sinusoidal symmetric perturbation to a $220 \, \text{nm}$-thick suspended silicon slab thus has an overall enhancement of a factor $10.2$ with respect to grating coupler with air below and a factor $12.5$ compared to a typical silicon-on-insulator grating coupler.

\subsection{Thickness dependence of scattering rate}

\begin{figure}[ht]
\hbox{\vspace{-5.3cm}\includegraphics[width=\linewidth,trim=3.7cm 9cm 3.3cm 9.0cm,clip]{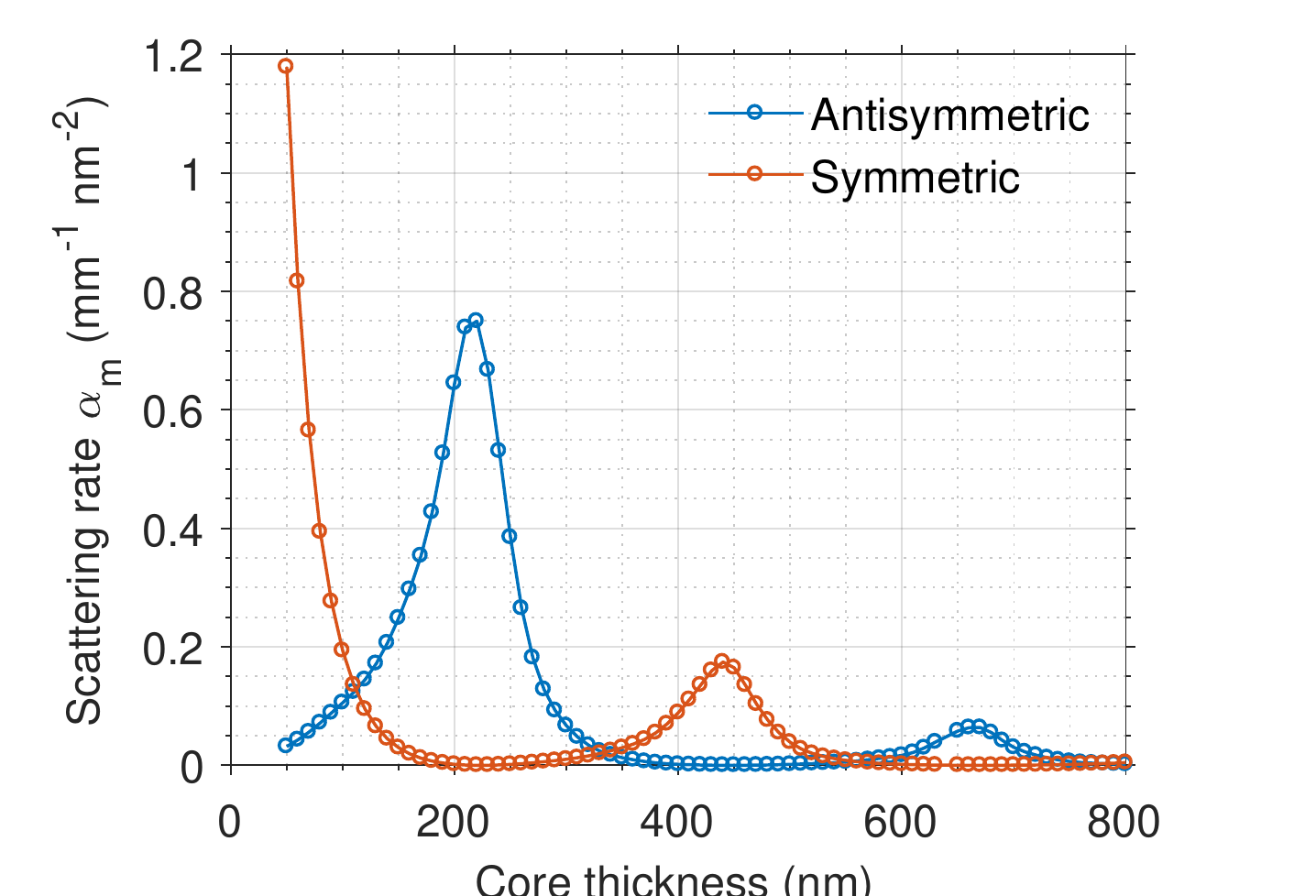}}
\hbox{\hspace{2.5cm}\includegraphics[width=0.2\linewidth,trim=5.2cm 24.0cm 14.0cm 1.5cm,clip]{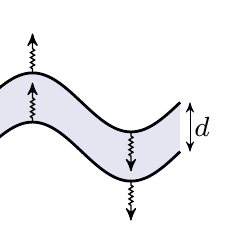}\vspace{0.45cm}}
\hbox{\hspace{4.05cm}\includegraphics[width=0.2\linewidth,trim=5.2cm 24.0cm 14.0cm 1.5cm,clip]{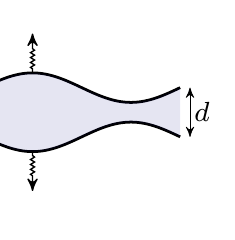}\vspace{1cm}}

\caption{\textbf{The scattering rate oscillates as a function of waveguide thickness.} The scattering rate $\alpham$ caused by the antisymmetric sinusoidal perturbation is maximal at thicknesses $d = \left(l +\frac{1}{2}\right)\lambda_{\text{Si}}$ with $\lambda_{\text{Si}} = \frac{\lambda}{n_{\text{Si}}} = 443 \, \text{nm}$ the optical wavelength in silicon and $l$ a positive integer, while it vanishes at $d = l\lambda_{\text{Si}}$. A symmetric sinusoidal perturbation yields the opposite behavior. This behavior is caused by interference between the scattered field from the top surface and that scattered from the bottom surface. As the waveguide core becomes thicker, the structure is less sensitive to either type of perturbation. One simulation point just above a thickness of $600 \, \text{nm}$ was removed manually as the script had selected the wrong optical mode. We computed both curves for a perturbation $u = 10 \, \text{nm}$.}
\label{fig:thickness_sweep}
\end{figure}

Next, we compute the scattering rate as a function of waveguide thickness for sinusoidal symmetric and antisymmetric perturbations (Fig. \ref{fig:thickness_sweep}) for the same parameters as in subsection \ref{subsec:comparison}.
We find the scattering rate $\alpha$ for the antisymmetric perturbation to be maximal when
\begin{equation}
d = \left(l +\frac{1}{2}\right)\lambda_{\text{Si}}
\end{equation}
with $d$ the core thickness, $l$ an integer and $\lambda_{\text{Si}} = \frac{\lambda}{n_{\text{Si}}} = 443 \, \text{nm}$ the optical wavelength in silicon. The scattering vanishes when
\begin{equation}
d = l \lambda_{\text{Si}}
\end{equation}
while the core with a symmetric perturbation exhibits the reverse behavior. The effect arises from interference between the scattered fields generated by the top and bottom slab surfaces. For an antisymmetric perturbation, there is another $180^{\circ}$ shift in the phasing of the top and bottom scatterers. Therefore, constructive interference occurs when the thickness is a multiple of a wavelength plus an additional half-wavelength to compensate for the $180^{\circ}$ phasing of the sources. Interestingly, the global maximum in the scattering rate of an antisymmetric perturbation -- similar to that of a Lamb-type mechanical field -- occurs exactly at a core thickness of $d = 220 \, \text{nm}$. In addition, this antisymmetric sinusoidal perturbation offers the strongest scattering rates of all the perturbation types (table \ref{table:comparison}). Hence antisymmetric Lamb-like flexural mechanical fields propagating along a $220 \, \text{nm}$-thick suspended silicon waveguide are ideal excitations for coupling guided and free-space optical fields.

\subsection{Photoelasticity}
\label{subsec:photoelasticity}
Even in absence of a geometric boundary perturbation, a propagating mechanical wave generates an inohomogeneous strain profile that couples guided optical fields to radiating fields. This is termed the photoelastic contribution $\alpha_{\text{pe}}$ to the total scattering rate $\alpha_{\text{tot}}$. The scattering rates reported above includes only the boundary-induced scattering $\alpha_{\text{mb}}$. Generally speaking, the two contributions may be of similar size and thus interfere with one another -- either enhancing the total scattering rate or potentially completely canceling it \cite{Kittlaus2015,VanLaer2015b,Rakich2012}. However, in a simulation without the boundary perturbation our eigenfrequency model predicts a photoelastic scattering rate of $\alpha_{\text{pe}} = 1.4 \cdot 10^{-2}/(\text{mm} \, \text{nm}^{2})$ -- nearly two orders of magnitude smaller than  $\alpha_{\text{mb}} = 0.74/(\text{mm} \, \text{nm}^{2})$ (table \ref{table:comparison}) for a suspended $220 \, \text{nm}$-thick silicon slab. Thus the photoelastic component of the scattering is weak in the considered geometry: even in case of completely destructive or constructive interference the total scattering rate
\begin{align}
\frac{\alpha_{\text{tot}}}{\alpha_{\text{mb}}} &= \left(1 \mp  \sqrt{\frac{\alpha_{\text{pe}}}{\alpha_{\text{mb}}}}\right)^2 = 1 \substack{+0.29 \\ -0.26}
\end{align}
would change by less than $30\%$. Next, we simulate the combined scattering rate resulting from interference between the moving-boundary and photoelastic scattering. We find that the two effects interfere constructively such that $\alpha_{\text{tot}} = 0.96/(\text{mm} \, \text{nm}^{2})$ -- about $30\%$ larger than $\alpha_{\text{mb}}$.

In this computation of $\alpha_{\text{pe}}$ and $\alpha_{\text{tot}}$ we implemented the photoelasticity as an anisotropic refractive index profile with components
\begin{equation}
n_{\text{core}} =
\begin{pmatrix}
    n_{\text{Si}} + \Delta n_{xx}      & 0 & \Delta n_{xz}  \\
    0       & n_{\text{Si}} + \Delta n_{yy} & 0 \\
    \Delta n_{xz} & 0 & n_{\text{Si}} + \Delta n_{zz}  \\
\end{pmatrix}
\end{equation}
where the index variations $\Delta n$ are given by
\begin{align}
\Delta n_{xx} &= -\frac{1}{2} n^{3}_{\text{Si}} p_{12} S_{zz} \\
\Delta n_{yy} &= -\frac{1}{2} n^{3}_{\text{Si}} p_{12} S_{zz} \\ 
\Delta n_{zz} &= -\frac{1}{2} n^{3}_{\text{Si}} p_{11} S_{zz} \\
\Delta n_{xz} &= - n^{3}_{\text{Si}} p_{44} S_{xz}
\end{align}
with $(p_{11},p_{12},p_{44}) = (-0.09,0.017,-0.051)$ the photoelastic tensor of silicon assuming waveguide orientation along a $\langle 100 \rangle$ axis. The waveguide orientation has a minor effect on the effective photoelastic components in silicon. The strain components are given by
\begin{align}
S_{zz} & =\partial_{z} u_{z}= -K^{2} x u\sin{Kz} \\
 S_{xz} &= \frac{1}{2}\left(\partial_{z}u_{x} + \partial_{x}u_{z}\right) \\
\notag &= S_{zx} = 0
\end{align}
where $K$ is the mechanical wavevector, $u$ the maximum mechanical perturbation and $(u_{x},u_{z}) = u (\sin{Kz},-Kx\cos{Kz})$ a snapshot of the displacement field of an antisymmetric Lamb-wave propagating along a thin slab opposite to the $z$-direction. The origin of the transverse coordinate $x$ is taken to be in the center of the slab, such that $|x| = d/2$ corresponds to the silicon/air interfaces with $d$ the slab thickness. This analytical Lamb-wave solution assumes $Kd \ll 1$. Finite-element simulations showed that the actual mechanical field around $Kd \approx 1$ is still captured well by this analytical approximation. In these simulations we swept $u$ and then obtained $\alpha_{\text{pe}}$ and $\alpha_{\text{tot}}$ from a fit to $u^{2}$ as in Fig. \ref{fig:verification}.

In general a full 3D simulation of the combined effects of moving boundaries and photoelasticity must be developed and is possible in our eigenfrequency approach. We have however limited our current 3D simulations to the moving-boundary effect given the expected weakness of photoelasticity in this system. We suspect that this weakness is caused by the distributed nature of the photoelastic scattering, leading to destructive interferences in the outgoing radiation similar to Fig. \ref{fig:thickness_sweep}.

\section{Perturbation theory for optomechanical antennas}
\label{sec:perturbationtheory}

In cavity optomechanics and Brillouin scattering, the optomechanical interaction is described perturbatively by expanding the permittivity in terms of the mechanical deformation
\begin{equation}
\perm \rightarrow \perm + \delta_u\perm \cdot \vecb{u} + \mathcal{O}\prens{\vecb{u}^2}.
\end{equation}
We can take the same approach to describe scattering out of a waveguide.
After a Fourier transform \( \partial_t \rightarrow -i\omega\), Maxwell's equations for the electric field reduce to
\begin{equation}
\prens{\nabla\times\nabla\times - \mu\perm\omega^2}\E = i\omega\mu \vecb{J}
\end{equation}
where \(\vecb{J}\) is a current density.
We are interested in a current-free waveguide with a mechanically perturbed permittivty
\begin{equation} \underbrace{\prens{\nabla\times\nabla\times - \mu\perm\omega^2}}_{\Large \mathbf{\Theta}}\E = \mu\omega^2\prens{\delta_u\perm\cdot\vecb{u}}\E.
\end{equation}

We first solve the uncoupled equations -- Maxwell's and the theory of elasticity -- for the optical and mechanical modes of the waveguide. The unperturbed electric field satisfies \(\mathbf{\Theta}\E = 0\), which can be rewritten as a generalized eigenvalue problem of the form $\mathbf{A}\E = \omega^2\mathbf{B}\E$ with Hermitian operators $\mathbf{A}$ and $\mathbf{B}$.
Expanding $\E \rightarrow \E + \E_\trm{r} +\dots$ and $\omega^2 \rightarrow \omega^2 + \omega^2_1 + \dots$ the perturbed field is given by
\begin{equation}
\mathbf{\Theta}\E_{\text{r}} = \mu\perm\omega_1^2\E + \mu\omega^2\prens{\delta_u\perm\cdot\vecb{u}}\E \end{equation} 
to first order in \(\vecb{u}\). Since \(\mathbf{\Theta}\) is Hermitian, \(\mathbf{\Theta}\ket{\E}=0\) implies that $\bra{\mathbf{E}}\mathbf{\Theta}\ket{\mathbf{E}_\mathbf{r}}=0$.  
Taking the inner product with $\ket{\E}$ on both sides yields the first order correction to the eigenvalue
\begin{equation}
\omega_1 = -\frac{\omega}{2} \frac{\bra{\E}\delta_u\perm\cdot\vecb{u}\ket{\E}}{\bra{\E}\perm\ket{\E}}
\end{equation}
where we've adopted Dirac notation and substituted \(\omega_1^2 \rightarrow 2\omega\omega_1\).  
For proper choice of inner product and normalization of $\vecb{u}$,  $\omega_1$ becomes the coupling rate of cavity optomechanics $g_0$ or Brillouin scattering $\tilde{g}_0$ \cite{Johnson2002,Aspelmeyer2014,VanLaer2016}. 

We can express the outgoing field fully in terms of \(\mathbf{\Theta}\) and the unperturbed fields 
\begin{equation}
\mathbf{\Theta}\E_{\text{r}} = \mu\omega^2\prens{\delta_u\perm\cdot\vecb{u}}\E.
\end{equation}
The optomechanical interaction above can be expressed in terms of a current 
\begin{equation}
\mathbf{J_\trm{om}} = -i\omega\prens{\delta_u\perm\cdot\vecb{u}}\E
\end{equation}
allowing us to rewrite the remaining inhomogeneous equations simply as
\begin{equation}
\mathbf{\Theta}\E_{\text{r}} = i\omega \mu\mathbf{J}_\trm{om}.
\label{eq:appendixAx=b}
\end{equation}

\subsection{The moving-boundary and photoelastic effect}

Next we present an explicit expression for the nonlinear polarization current $\mathbf{J}_\trm{om}$. Mechanical waves give rise to changes in the permittivity $\perm$ which can be expressed to first order in $\vecb{u}$ in terms of body (\ie photoelastic effect~\cite{Dixon1967}) and boundary contributions. The latter requires careful handling to manage discontinuities in the field at boundaries of dielectrics treated by Johnson \emph{et al.}~\cite{Johnson2002,Johnson2005}.
This variation in the permittivity $\delta_u\perm\cdot\vecb{u}$, which is familiar in the fields of cavity optomechanics and Brillouin scattering, is here a tensor which acts on $\E$ to give the polarization currents $\vecb{J}_\trm{om}$ above. 

Consider a domain \(\Omega\) with boundary \(\partial\Omega\) deformed by \(\vecb{u}\).
The normal \(\vecb{n}\) points out of the domain such that for positive \(\vecb{u}\cdot\vecb{n}\) the permittivity of a region in the neighborhood of the boundary changes by \(\Delta\perm \equiv            \perm_\trm{in} - \perm_\trm{out}\).
The main trick in forming a well-defined expression for the radiation   pressure on the boundary is to avoid field discontinuities by replacing the   component of \(\vecb{E}\) normal to \(\partial\Omega\) with the electric displacement field \(\vecb{D}\).
The boundary contribution then becomes
\[ \delta_u\perm_\trm{rp}\cdot\vecb{u} =                                \prens{\vecb{u}\cdot\vecb{n}}\delta_{\partial\Omega}\prens{\Delta\perm\vecb{\Pi}_\shortparallel - \perm\Delta\perm^{-1}\perm\vecb{\Pi}_\perp}\]
which is expressed in terms of the tensors \(\vecb{\Pi}_\shortparallel\) and     \(\vecb{\Pi}_\perp = \mathbb{I}-\vecb{\Pi}_\shortparallel\) which project the electric field into the  plane of the dielectric interface or along the normal \(\vecb{n}\), respectively.
Here \(\Delta\perm^{-1} \equiv \perm^{-1}_\trm{in} - \perm^{-           1}_\trm{out}\) and the delta function on the boundary                   \(\delta_{\partial\Omega}\) renders                                     \(\prens{\delta_u\perm\cdot\vecb{u}}\vecb{E}\) into a surface current   on \(\partial\Omega\).

Suppose the normal is oriented in Cartesian coordinates along \(\vecb{z}\) and the dielectric boundary is at \(z = 0\).
Then
\[ \prens{\delta_u\perm_\trm{rp}\cdot\vecb{u}}\vecb{E} =                u_z\delta\prens{z}\begin{bmatrix}\Delta\perm E_x \\ \Delta\perm E_y \\ -\perm\Delta\perm^{-1}D_z\end{bmatrix}.
\label{eq:rpExplicit}\]
The $\vecb{z}$ component of the expression above poses some difficulty as $\perm$ is discontinuous as $z=0$.  This is discussed in the next section.

	Although in this work estimates of the photoelastic effect justified    dropping it from our calculations, we give its contribution to the      variation of \(\perm\) below for completeness. The photoelastic  tensor in common use is defined such that a strain \(\vecb{S}\) causes  \(\perm\Delta\perm^{-1}_{ij} = \prens{\vecb{p}\vecb{S}}_{ij} =        p_{ijkl}S_{kl}\).
	To first order in \(\vecb{u}\),
	\[ \delta_u\perm_\trm{pe}\cdot\vecb{u} = -\perm                         \frac{\vecb{p}\vecb{S}}{\perm} \perm. \]

\subsection{Implementing the boundary contribution to $J_\trm{om}$}

The boundary contribution to the scattering process is delta-distributed across the boundary.  
In the plane of a dielectric interface -- along the  $\vecb{x}$ and $\vecb{y}$ axis in equation~(\ref{eq:rpExplicit}) -- the perturbation behaves like a surface current giving rise to discontinuities in the magnetic field 
\[ \vecb{n}\times \Delta\vecb{H} = \vecb{J}_\shortparallel. \]

The $J_\perp$ component of the surface current isn't implemented as a set of boundary conditions. 
Instead $J_\perp$ is taken to be a uniform volume current density of finite thickness $t_J = 5~\nm$ just inside the boundaries of our silicon waveguides, normalized by the thickness so as to converge to a delta function the limit $t_J \rightarrow 0$.

The expression for $J_\perp$ is the product of a step and a delta function at the boundary, and volume approximations of the delta function therefore require some choice of $\perm$ in the $\vecb{z}$ component of equation~(\ref{eq:rpExplicit}).  Although $J_\perp$ is discontinuous, the power sourced into the field $-\E\cdot \vecb{J}_\trm{om}$ \emph{is} continuous, making the radiated field robust to the exact distribution used for $J_\perp$~\cite{Johnson2005}.  

We check our implementation of $J_\perp$ by computing the OM radiation of the fundamental TM modes of $220~\trm{nm}$, $340~\trm{nm}$, and $440~\trm{nm}$ silicon slab waveguides both perturbatively and nonperturbatively.

\begin{figure}[ht]
\includegraphics[width=\linewidth]{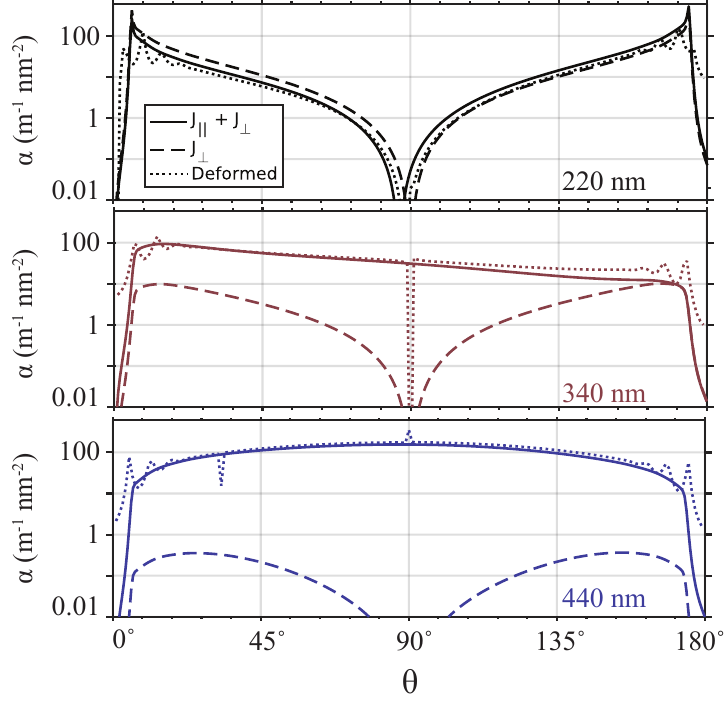}
\caption{The scattering rates $\alpha$ are computed for, from top to bottom, a 220~nm, 340~nm, and 440~nm slab with and without the volume current used to implement the $J_\perp\propto D_z$ component of the boundary interaction.  For the 220~nm slab, the $J_\perp$ term dominates and our perturbative (dashed) and nonperturbative (dotted) calculations agree well.}
\label{fig:TMSlab}
\end{figure}

\section{Geometric disorder and dephasing}
\label{sec:geometricdisorder}
Geometric disorder has been studied extensively in nanophotonic circuits \cite{Selvaraja2011,Melati2014,Elson1995,Fursenko2012,Lee2000,Yang2015a} to understand optical propagation losses. It has also been investigated in the context of Brillouin scattering -- where it leads to broadening of the mechanical resonance \cite{Wolff2016a,VanLaer2015,VanLaer2015b,Kittlaus2015}. In both cases the standard deviation of geometric disorder was generally estimated at $\sigma \approx 1 \, \text{nm}$ with the largest disorder occurring near etched sidewalls. In studies focusing on optical scattering losses, the coherence lengths $\xi$ were found to be below $100 \, \mu\text{m}$ \cite{Fursenko2012,Lee2000}. Such short coherence lengths are not consistent with (1) optically measured wafer-scale geometric disorder \cite{Selvaraja2011}, (2) measured millimeter-scale optical dephasing lengths \cite{Yang2015a} nor (3) with the Brillouin resonance broadening observed in silicon waveguides \cite{VanLaer2015b,Kittlaus2015,Wolff2016a}. Therefore, we suspect the spatial correlator given in equation (\ref{eq:noiseCorrelations}) to contain at least two terms: (1) a fast-disorder term with a coherence length of about $50 \, \text{nm}$ and (2) a slow-disorder term with a coherence length of about $50 \, \mu\text{m}$. The fast roughness mainly determines optical radiation loss and backscatter -- both of which require large roughness momentum -- while the slow drift mainly determines the dephasing as it builds up over many wavelengths. We expect slow disorder to be the main hurdle for the proposed device and thus estimate the various sensitivities in table \ref{table:dephasing} using $\xi \approx 50 \, \mu\text{m}$.

In the main text, we provided the results of our analysis of dephasing. Here we discuss the derivations in detail. There are eight cases to be investigated: the out- vs. incoupling, anti-Stokes vs. Stokes and counter- vs. copropagating cases could each be combined. Four of these cases suffer from low efficiencies due to a large phase-mismatch. The remaining four cases are (1) outcoupling by anti-Stokes scattering between counter-propagating guided waves, (2) outcoupling by Stokes scattering between copropagating guided waves, (3) incoupling by Stokes scattering between copropagating guided waves and (4) incoupling by anti-Stokes scattering between counter-propagating guided waves. Cases (3) and (4) are the time-reversed versions of cases (1) and (2) respectively. Thus they have identical properties and we limit ourselves to cases (1) and (2) in the following.

Case (1) is attractive as it allows for optical and mechanical excitation from opposite sides of the array. In anti-Stokes scattering phase errors of the guided optical and the guided mechanical wave add, yielding a total phase error of
\begin{equation}
\delta \varphi(z) = \int_0^{z}dz' \, \delta \beta(z') - \int_L^{z}dz' \, \delta K(z') 
\end{equation}
The optical phase error grows forwards while the mechanical phase error grows backwards. The variance of the phase error is
\begin{align}
\notag \langle \delta\varphi^{2}(z) \rangle &= \int_{0}^{z} \int_{0}^{z} dz' dz'' \, \langle \delta \beta(z') \delta \beta(z'') \rangle \\
\notag & + \int_{L}^{z} \int_{L}^{z} dz' dz'' \langle \delta K(z') \delta K(z'')\rangle \\
& + 2 \int_{0}^{z}\int_{z}^{L} dz'dz'' \, \langle \delta\beta(z')\delta K(z'')\rangle
\label{eq:phasevariance}
\end{align}
To compute these terms, we expand $\delta \beta = \sum_l \partial_l \beta \delta X_l$ and $\delta K = \sum_l \partial_l K \delta X_l$ and insert the spatial correlator of equation (\ref{eq:noiseCorrelations}). Next, there are two ways to proceed with the first two terms. Either one makes direct use of the integral
\begin{align}
\notag \int_{0}^{z} \int_{0}^{z} \text{d}z'' \text{d}z' \, e^{-|\Delta z|/\xi_l} &= 2\xi_l \left(z - \xi_l  + \xi_l e^{-z/\xi_l} \right) \\
&\overset{z \gg \xi_l}{\approx} 2\xi_l z
\end{align}
with $\Delta z = z' - z''$ and similarly for the $\delta K$-term. In an alternative approach, we define the power spectral density $S_{\beta\beta}[G]$ of $\delta \beta$ as 
\begin{equation}
S_{\beta \beta}[G] = \int_{-\infty}^{+\infty} dz \, e^{iGz} \langle \delta \beta(z)\delta \beta(0)\rangle
\end{equation}
Using equation (\ref{eq:noiseCorrelations}) this leads to
\begin{equation}
S_{\beta \beta}[G] = \sum_l \frac{2(\partial_l \beta \sigma_l)^{2} \xi_l}{1 + \left(G \xi_l\right)^{2}}
\end{equation}
For $z \gg \xi_l$ and $L - z \gg \xi_l$ the $\delta\beta \delta K$ cross-term in equation (\ref{eq:phasevariance}) is negligible such that
\begin{equation}
\left\langle\delta\varphi^2\prens{z}\right\rangle = S_{\beta\beta}[0] z + S_{KK}[0](L-z)
\end{equation}

In case (2), Stokes scattering implies that the guided optical and guided mechanical phases subtract such that
\begin{equation}
\delta \varphi(z) = \int_{0}^{z} dz' \, \delta k_{\shortparallel}
\end{equation}
with $k_{\shortparallel} = \beta - K$ so one can show as in the above that
\begin{equation}
\langle \delta\varphi^{2}(z) \rangle = S_{k_{||}k_{||}}[0]z
\end{equation}
The thermal dephasing lengths in the main text were derived along similar lines but with constant $\delta \beta$ and $\delta K$ along the antennas.

\section{Mechanical losses}

Material losses cause mechanical waves to decay at a rate $\gamma = \Omega/(Q_\trm{m} v_\trm{m})$ where $Q_\trm{m}$ is the mechanical quality factor and $v_\trm{m}$ is the mechanical group velocity. 
At room temperature, nonlinear phonon processes limit mechanical $Q$s of silicon resonators to $Q\approx 10^4$~\cite{Ghaffari2013}.
Waves in the OMA of Section~\ref{sec:OMAsforsiliconphotonics} at $K = 2\pi\times 1.2~\um^{-1}$ which scatters light at $\theta = 60^\circ$ have a frequency $\Omega = 2\pi\times 3.42~\trm{GHz}$ and group velocity $v_\trm{m} = 4135 \, \text{m/s}$.
Assuming a quality factor $Q_\trm{m} =  3\times 10^3$, the resulting decay length is  $\gamma^{-1} = 580~\um$.  

\begin{figure}[ht]
	\includegraphics[width=\linewidth]{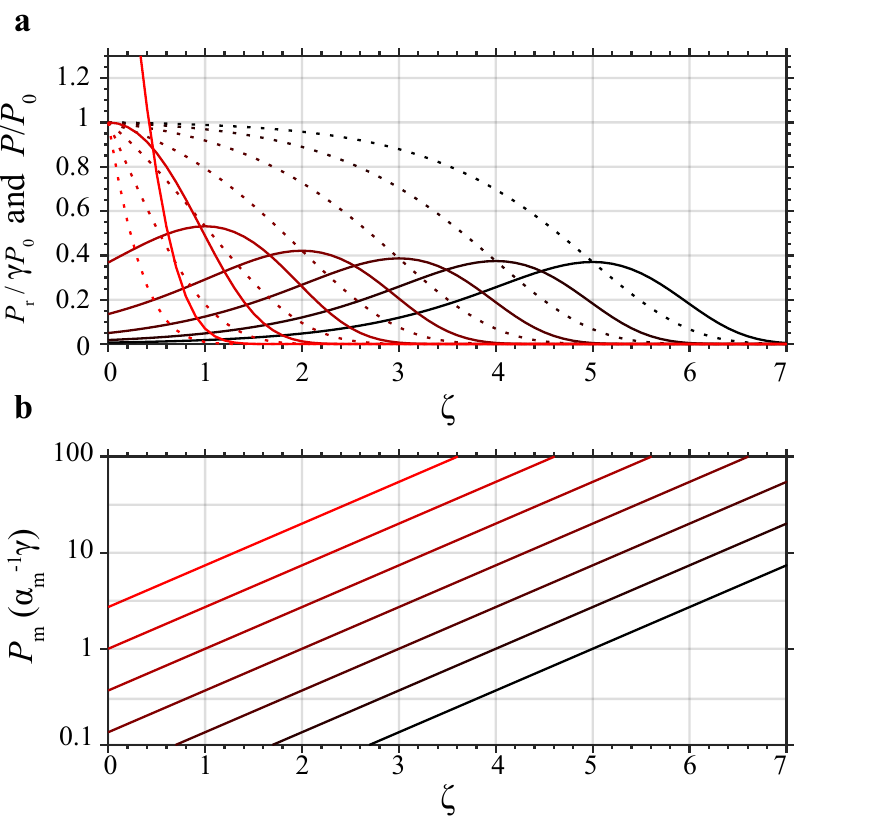}
    \caption{\textbf{Mechanical attenuation and the effective aperture.} Mechanical losses alter the exponential envelope of an ideal OMA, limiting the effective aperture and increasing the mechanical power needed to achieve it. \textbf{a}  The solid curves show power radiated per unit length across an antenna for values of $\zeta_\trm{max} = -\log a$ ranging from -1 to 5 (from red-to-black in even steps).  Large scatterings rates $a$ results in approximately exponential radiation patterns as in the lossless case.  As $a$ decreases, the profile shifts toward higher $z$ with a constant FWHM of $\Delta\zeta = 2.45$.  The dotted curves show the fraction of the guided light remaining in the waveguide, plotted on the same axes.  \textbf{b} The mechanical power needed to achieve a particular radiation pattern is plotted against the antenna's length $L$ on the same abscissa as \textbf{a}. 
    }
	\label{fig:attenuationCurves}
\end{figure}

Attenuation of the mechanical waves modifies the optical scattering rate $\alpha$ from the antenna and thereby the radiation pattern of an OMA.
Consider the counter-propagating optical and mechanical waves of the anti-Stokes process for which 
\begin{align}
\partial_z \mathcal{P} &= -\alpham \mathcal{P}_\trm{m}\mathcal{P} \\
\partial_z \mathcal{P}_\trm{m} &= \gamma\mathcal{P}_\trm{m}
\end{align}
where $\alpham$ is a power-normalized scattering rate as plotted in Fig.~\ref{fig:OM_scattering}c, not to be confused with the displacement-normalized rate represented by $\alpham$ in the rest of the text. 
In the above equations we assume the mechanical drive $\mathcal{P}_\trm{m}$ is undepleted by the scattering process, a reasonable assumption since the phonon flux for a 1~mW drive is larger than a 1~mW optical guided wave by a factor of $\omega/\Omega \approx 10^{5}$.
The optical power is $\mathcal{P}_0$ at the beginning of the antenna where $z = 0$ and the mechanical power is $\mathcal{P}_{\trm{m}\,0}$ at the end of the antenna where $z = L$.
We can solve these equations given the boundary conditions above to find the optical power along the antenna
\begin{equation}
	\frac{\mathcal{P}\prens{z}}{\mathcal{P}_0} =  e^{-a \prens{e^{\zeta} - 1}}
\end{equation}
where we've introduced two dimensionless parameters: the local scattering rate at the beginning of the antenna $a = -\alpham\mathcal{P}_{\trm{m}\, 0} e^{-\gamma L}/\gamma$ and the distance along the antenna $\zeta = \gamma z$.
The power radiated from the waveguide is computed from the guided power by taking the derivative $\mathcal{P}_\trm{r} = -\partial_z \mathcal{P}$ yielding
\begin{equation}
	\frac{\mathcal{P}_\trm{r}\prens{\zeta}}{\gamma\mathcal{P}_0} = a \exp\prens{\zeta - a\prens{e^\zeta - 1}}.
\end{equation}
Maximal scattering occurs at $\zeta_\trm{max} = -\log a$.
When the optical scattering rate at $z = 0$ is large compared to the mechanical attenuation rate $\gamma$ such that $\zeta_\trm{max} \ll 0$,  the power radiated $\mathcal{P}_\trm{r}$ is well described by equation~(\ref{eq:opticalPowerAttenuation}) and attenuation can be ignored.
As $a$ is decreased perhaps by lowering $\mathcal{P}_{\trm{m}\, 0}$, the maximum shifts right $\zeta_\trm{max}$ and when sufficiently far from the origin and for sufficiently long antennas the resulting radiation pattern has a FWHM of $\Delta \zeta = 2.45$.
The 1/$e$ width is $3$ and the 1/$e^2$ width is 4.45.
Figure~\ref{fig:attenuationCurves} shows the mechanical power necessary to achieve a particular radiation pattern for an antenna of a particular length.

\section{Derivation of antenna properties}
\label{sec:antennapropertiesderivation}
\begin{figure}[ht]
\includegraphics[width=\linewidth]{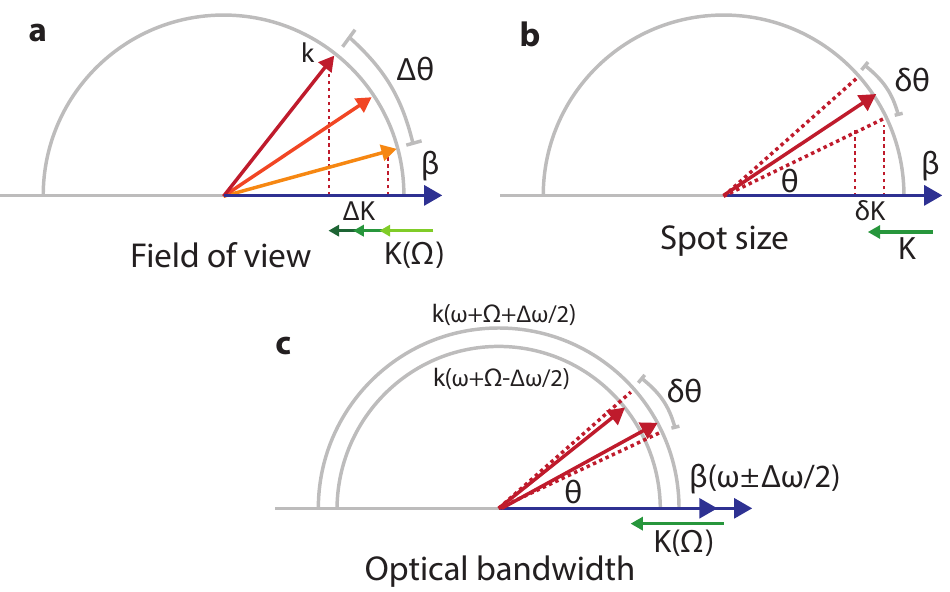}
\caption{\textbf{Optomechanical antenna properties.} \textbf{a}, The field of view $\Delta\theta$ is the range of angles that can be reached by tuning the mechanical frequency $\Omega$ at fixed optical frequency $\omega$. \textbf{b}, The spot size $\delta\theta$ is the angular width -- set by the wavevector uncertainty $\delta K$ -- of the scattered optical beam in the far field. \textbf{c}, The optical bandwidth at each spot $\Delta\omega$ is the amount the optical frequency $\omega$ can be shifted before the beam angle shifts by more than the spot size $\delta \theta$ at fixed mechanical frequency $\Omega$. Since both the outgoing wavevector $k$ and the guided wavevector $\beta$ change with frequency, the optical bandwidth at each spot $\Delta \omega$ is set by the walk-off between the guided and radiating optical fields.}
\label{fig:FOMs}
\end{figure}

Here we provide derivations of the antenna properties presented in the main text. Some of the properties are illustrated in Fig. \ref{fig:FOMs}.

\subsection{Field of view}

The field of view $\Delta \theta$ is the range of angles than can be reached by sweeping the mechanical frequency. It is set by the bandwidth of the electromechanical transducer, the sensitivity of the mechanical wavevector to frequency as well as the sensitivity of radiation wavevector to mechanical wavevector. In particular, subtracting the phase-matching conditions (\ref{eq:phasematching}) at two different mechanical frequencies leads to
\begin{equation}
k(\omega_\trm{r} + \Delta \Omega) \cos(\theta + \Delta\theta) - k(\omega_\trm{r})\cos(\theta) = -\Delta K
\end{equation}
with $\Delta \Omega= v_{\text{m}} \Delta K$ the bandwidth of the electromechanical transducer. This determines $\Delta \theta$ in general, while for small $\Delta \theta$ we get
\begin{align}
\Delta \theta &= \frac{\Delta \Omega}{k \sin(\theta)}\left(\frac{1}{v_{\text{m}}} + \frac{\cos(\theta)}{c} \right) \\
&\approx \frac{\Delta \Omega}{k \sin(\theta) v_{\text{m}}} = \frac{\Delta \Omega}{2\pi}\frac{\lambda}{\sin(\theta) v_{\text{m}}}
\end{align}
with $\lambda = 2\pi/k$ the radiation's wavelength. Here we used $c \gg v_{\text{m}}$.

\subsection{Spot size}
\label{sec:spotsize}
The spot size $\delta \theta$ is the angular width of the scattered optical beam in the far field. It is set by the wavevector uncertainty $\delta K = \frac{2\pi}{L_{\text{eff}}}$ corresponding to the finite aperture:
\begin{align}
\label{eq:spotsize}
k \, \delta\cos(\theta) &\approx k\sin(\theta)\delta \theta \\ \notag
&= \delta K = \frac{2\pi}{L_\textrm{eff}}
\end{align}
where we dropped a minus sign. Therefore,
\begin{equation}
\delta \theta = \frac{\lambda}{\sin(\theta)L_{\text{eff}}}
\end{equation}

\subsection{Number of resolvable spots}
The number of resolvable spots $N_\theta$ is the ratio between the field of view $\Delta \theta$ and the spot size $\delta \theta$. Neglecting the frequency-dependence of $k$ allows us to express $N_\theta = \Delta K / \delta k$ only in terms of the mechanical properties of the optomechanical antenna. We find that 
\begin{equation}
N_\theta = \frac{\Delta K}{\delta k} = \frac{\Delta \theta}{\delta \theta} = \frac{\Delta\Omega}{2\pi}\,\tau_\textrm{m}.
\end{equation}
The number of resolvable spots is set by the product of the transducer bandwidth $\Delta\Omega$ and the mechanical transit time $\tau_\textrm{m} = L_\text{eff}/v_{\text{m}}$. With a large bandwidth transducer we have $\Delta K \rightarrow 2k$ and therefore $N_\theta \rightarrow 2\frac{L_\textrm{eff}}{\lambda}$ -- making the effective aperture length the ultimate limit on the number of resolvable spots.

\subsection{Bandwidth at each spot}
\label{subsec:opticalbandwidth}
The optical bandwidth is set by how much the optical frequency can be changed before the beam angle disperses more than the spot size. For fixed $K$ we differentiate the phase-matching condition (\ref{eq:phasematching}) and
\begin{equation}
\cos\theta\,\delta k + \underbrace{k \, \delta\cos\theta}_{= \delta K} = \delta\beta.
\end{equation}
We equate the second term, the angular variation from changing $\omega$, to the angular spread of the beam in section \ref{sec:spotsize} $\delta K = \frac{2\pi}{L_\textrm{eff}}.$ 
Relating $\delta\beta$ and $\delta k$ to $\Delta\omega$ by the guided and free-space optical dispersion relations we find
\begin{equation}
\frac{\Delta\omega}{2\pi} = \frac{1}{\tau - \tau_\trm{r}}
\end{equation}
with $\tau = L_{\text{eff}} n_{\text{g}}/c$ the transit time of the guided optical wave and $\tau_\trm{r} = L_{\text{eff}} \cos{\theta}/c$ the transit time of the radiation mode across the aperture and $n_{\text{g}}$ the group index of the guided optical wave. The optical bandwidth at each spot $\Delta\omega_{0}$ is thus set by the walk-off between the guided and the radiation mode.

A similar derivation as in the optical case above shows that the mechanical transit time determines the mechanical bandwidth at each spot
\begin{equation}
\frac{\Delta \Omega_{\text{m}}}{2\pi} = \frac{1}{\tau_{\text{m}}}
\end{equation}

\end{document}